\documentclass{article}

\usepackage{epsfig} 
\newlength{\abstractwidth} 
\renewcommand{\thefootnote}{\fnsymbol{footnote}} 
\renewcommand{\thanks}[1]{\footnote{#1}} 
\newcommand{\starttext}{ 
\setcounter{footnote}{0} 
\renewcommand{\thefootnote}{\arabic{footnote}}} 
\renewcommand{\theequation}{\thesection.\arabic{equation}} 
\newcommand{\bea}{\begin{eqnarray}} 
\newcommand{\eea}{\end{eqnarray}} 
\newcommand{\beq}[1]{\begin{equation} \label{#1}} 
\newcommand{\be}{\begin{equation}} 
\newcommand{\ee}{\end{equation}}

\def\12{{1 \over 2}}

\newcommand{\tab}{\hspace{5mm}}

\begin{document} 
\renewcommand{\theequation}{\thesection.\arabic{equation}} 
\begin{titlepage} 
\bigskip
\centerline{\Large \bf {An Extended Poincare Algebra for Linear Spinor Field Equations}}
\bigskip
\begin{center} 
{\large James Lindesay\footnote{
Permanent address, Department of Physics, Howard University, Washington, DC 20059}
} \\
\end{center}
\bigskip\bigskip 
\begin{abstract} 

When utilizing a cluster decomposible relativistic scattering formalism, it
is most convenient that the covariant field equations take on a linear form
with respect to the energy and momentum dispersion on the fields in the
manner given by the Dirac form for spin ${1 \over 2}$ systems.  The general spinor
formulation for arbitrary spins given in a previous paper is extended to include 
momentum operators.  Unitary quantum mechanical representations are
developed for these operators, and physical interpretations are suggested.

\end{abstract} 
\end{titlepage} 


\starttext \baselineskip=17.63pt \setcounter{footnote}{0} 
\setcounter{equation}{0} 
\section{ Introduction} 
\tab 
The Dirac equation\cite{Dirac} utilizes a matrix algebra to construct a linear operator
relationship between the energy and the momentum.  Such a linear dispersion
relationship is particularly useful for constructing manifestly cluster decomposible
non-perturbative scattering formalisms\cite{LMNP}\cite{AKLN}\cite{Compton}. 
Expectation values of the matrices can be related to physical fluxes, but the matrices
themselves commute with space-time translation generators.  In a previous paper\cite{jlspinor},
a general operator representation of an extended Lorentz group which
yielded the Dirac matrix representation for spin ${1 \over 2}$ systems was
developed.  If the transformation is internal, this previous development is sufficient
to develop quantum mechanical amplitudes that can represent general particle systems. 
In this paper we will extend the group to include the generators
for infinitesimal space-time translations, allowing global transformations in these
group parameters.

The inclusion of the expected commutation relations between energy-momenta,
angular momenta, and boost operators will require non-vanishing commutation
relations between energy-momenta and $\Gamma^\mu$ operators.  In addition,
group closure requires the addition of at least one additional operator to the algebra. 
The interpretations of this extended Poincare algebra will be explored briefly
subsequent to its development.
 
The finite dimensional representations of the extended Poincare group will be
constructed with use of the little group and commplimentary group of transformations on
the standard state vectors in the construction of particle-like unitary
representations for spin states.  In addition,
a single-particle wave equation can be developed for configuration space
eigenstates of the operator $\hat{\Gamma}^\mu \: \hat{P}_\mu$,
\be
-\mathbf{\Gamma}^\mu \: i \partial_\mu \psi (x) \: = \: \lambda \: \psi (x)
\ee
which implies that
\be
\partial_\mu \left [ \overline{\psi (x)} \: \mathbf{\Gamma}^\mu \: \psi (x) \right ] \: = \: 0.
\label{current}
\ee
The conserved current defined in Equation \ref{current} need not be the
probability current, since the spinor metric is generally not directly related to the
$\Gamma$ matrices.  However, scattering equations can be
developed to express the evolution of the physical parameter represented by this operator.

Finally, we will see that transformations by the new extended group parameters
mixes what would be the mass parameter for traditional particle states with
the new transverse mass parameter generated by the additional operator.  We
will briefly examine the behavior of states transformed by these extended group
operators.

\setcounter{equation}{0} 
\section{Extended Lorentz Group} 

\tab 
As explained in the introduction, we consider it
advantageous to extend the previously developed extended Lorentz 
group to include the usual Poincare algebra, but generally require that the form
$\hat{\Gamma} ^\mu \hat{P}_\mu$ is a scalar operator.  In a previous
paper \cite{jlspinor}, spinor and matrix representations were developed for
the generators conjugate to the 10 group elements given by
3 rotations $\theta_k$, 3 Lorentz boosts $v_k$, and the 4 group parameters
associated with the generators $\hat{\Gamma}^\mu$ given by $\vec{\omega}$. 
For completeness, we will here briefly review those relationships.

\subsection{Extended Lorentz Group Commutation Relations} 
\tab
The commutation relationships between the generators of this extended
Lorentz group are given by
\bea
\left [ J_j \, , \, J_k \right] \: = \: i \, \epsilon_{j k m} \, J_m 
\label{JJeqn} \\
\left [ J_j \, , \, K_k \right] \: = \: i \, \epsilon_{j k m} \, K_m 
\label{JKeqn} \\
\left [ K_j \, , \, K_k \right] \: = \: -i \, \epsilon_{j k m} \, J_m 
\label{KKeqn} \\
\left [ \Gamma^0 \, , \, \Gamma^k \right] \: = \: i \, K_k 
\label{Gam0Gameqn} \\
\left [ \Gamma^0 \, , \, J_k \right] \: = \: 0 
\label{Gam0Jeqn} \\
\left [ \Gamma^0 \, , \, K_k \right] \: = \: -i \,  \Gamma^k 
\label{Gam0Keqn} \\
\left [ \Gamma^j \, , \, \Gamma^k \right] \: = \: -i \, \epsilon_{j k m} \, J_m 
\label{GamGameqn} \\
\left [ \Gamma^j \, , \, J_k \right] \: = \: i \, \epsilon_{j k m} \, \Gamma^m 
\label{GamJeqn} \\
\left [ \Gamma^j \, , \, K_k \right] \: = \: -i \, \delta_{j k} \, \Gamma^0 
\label{GamKeqn}
\eea

An extended Lorentz group Casimir operator can be constructed in the form
\be
C_\Lambda \: = \: \underline{J} \cdot \underline{J} \,-\, \underline{K} \cdot \underline{K}
\,+\, \Gamma^0 \, \Gamma^0 \,-\, \underline{\Gamma} \cdot \underline{\Gamma}.
\label{Casimir_Lam}
\ee

\subsection{Spinor Equations} 
\tab 
We found that the most elegant way to construct general finite
dimensional representations was by utilizing the formalism of
spinors.  Operations in the form of raising and lowering operators
are most conveniently expressed in terms of spinors. 
For convenience, define $\Delta_k ^{(\pm)}$ as follows:
\be
\Delta_k ^{(\pm)} \: \equiv \: \Gamma^k\, (\pm) \, i  K_k  .
\ee
We chose to construct eigenstates of $C_\Lambda$, $\Gamma^0$, $J^2$, and
$J_z$.  To develop a basis of states,
it was convenient to construct an operator which raises and lowers eigenvalues
of the operator $\Gamma^0$.  This operator is given by
\be
\Delta_J ^{(\pm)} \: \equiv \: \underline{J} \cdot \underline{\Delta}^{(\pm)}
\ee
These definitions allow us to express general states and operators in terms
of the spinors:
\be
\begin{array}{r}
\psi_{\gamma , M} ^{\Lambda , J} \: = \: A^{\Lambda J}
\sqrt{{(J-M)! \over (J+M)! \, (2J)!}} \,[x-y]^{\Lambda - J} \, 
\chi_+ ^{(+) \, M+\gamma} \, \chi_+ ^{(-) \, M-\gamma} \,\times \\
\\
\left .
\left[ {\partial \over \partial x} \,+\, {\partial \over \partial y}   \right] ^{J+M}
x^{J-\gamma} \, y^{J+\gamma} \right | _{
\begin{array}{l}
x=\chi_+ ^{(+)} \, \chi_- ^{(-)} \\ y=\chi_- ^{(+)} \, \chi_+ ^{(-)} 
\end{array} }
\end{array}
\ee

\bea
\hat{J}^2 \, \psi_{\gamma , M} ^{\Lambda , J} \: = \: 
J(J+1) \,\psi_{\gamma , M} ^{\Lambda , J} \quad , \quad &
\hat{C} \, \psi_{\gamma , M} ^{\Lambda , J} \: = \: 
2 \Lambda (\Lambda+2) \,\psi_{\gamma , M} ^{\Lambda , J} \\
\hat{J}_z \, \psi_{\gamma , M} ^{\Lambda , J} \: = \: 
M \,\psi_{\gamma , M} ^{\Lambda , J} \quad , \quad &
\hat{\Gamma}^0 \, \psi_{\gamma , M} ^{\Lambda , J} \: = \: 
\gamma \,\psi_{\gamma , M} ^{\Lambda , J} \\
\hat{J}_\pm \, \psi_{\gamma , M} ^{\Lambda , J} \: = \: 
\sqrt{(J \pm M + 1)(J \mp M)} \,\psi_{\gamma , M \pm 1 } ^{\Lambda , J} \quad , \quad &
\hat{\Delta}^{(\pm)} \, \psi_{\gamma , M} ^{\Lambda , J} \: = \: 
(\pm)\,(\Lambda+1)\,[J (\mp) \gamma] \,\psi_{\gamma \pm 1 , M} ^{\Lambda , J}
\label{op-spinor}
\eea

\subsection{Spinor metric}
\tab
Invariant amplitudes that can be interpreted as probability amplitudes
are defined using dual spinors, so that
under transformations the inner product is a scalar
\be
\begin{array}{c}
<\bar{\psi} | \phi > \: = \: <\bar{\psi'} | \phi'> \\
\psi_a ^\dagger g_{ab} \phi_b \: = \: 
\left( D_{ca} \psi_a  \right) ^\dagger g_{cd} 
\left( D_{db} \psi_b  \right)
\end{array}
\ee
The metric satisfies
\be
g_\gamma ^{\Lambda J} \: = \: (-)^{\Lambda-\gamma},
\ee
which will be associated with $\overline{\psi}$ in matrix element
evaluations.

\subsection{Matrix Representation for $\Lambda=1$}
\tab
As a specific example, the matrix representation corresponding to
$\Lambda=1$ will be explicitly demonstrated below:
\be
\begin{array}{c c}
\mathbf{\Gamma^0} \,=\, \left ( \begin{array}{r r r r}
0 & \underline{0} & \underline{0} & \underline{0} \\
\underline{0} & \mathbf{1} & \mathbf{0} & \mathbf{0} \\
\underline{0} & \mathbf{0} & \mathbf{0} & \mathbf{0} \\
\underline{0} & \mathbf{0} & \mathbf{0} & \mathbf{-1} 
\end{array} \right ) &
\mathbf{\Gamma^k} \,=\, {1 \over 2} \left ( \begin{array}{c c c c}
0 & 2\underline{v}_k ^T & \underline{0} & 2\underline{v}_k ^T \\
-\underline{v}_k & \mathbf{0} & \mathbf{J}_k & \mathbf{0} \\
\underline{0} & -2 \mathbf{J}_k & \mathbf{0} & 2 \mathbf{J}_k \\
-\underline{v}_k & \mathbf{0} & -\mathbf{J}_k & \mathbf{0}  
\end{array} \right )  \\ \\
\mathbf{J_k} \,=\, \left ( \begin{array}{c c c c}
0 & \underline{0} & \underline{0} & \underline{0} \\
\underline{0} & \mathbf{J}_k & \mathbf{0} & \mathbf{0} \\
\underline{0} & \mathbf{0} & \mathbf{J}_k & \mathbf{0} \\
\underline{0} & \mathbf{0} & \mathbf{0} & \mathbf{J}_k 
\end{array} \right ) &
\mathbf{K_k} \,=\, {1 \over 2} \left ( \begin{array}{c c c c}
0 & -2\underline{v}_k ^T & \underline{0} & 2\underline{v}_k ^T \\
-\underline{v}_k & \mathbf{0} & \mathbf{J}_k & \mathbf{0} \\
\underline{0} & 2 \mathbf{J}_k & \mathbf{0} & 2 \mathbf{J}_k \\
\underline{v}_k & \mathbf{0} & \mathbf{J}_k & \mathbf{0}  
\end{array} \right )  \\ \\
\mathbf{g} \,=\, \left ( \begin{array}{r r r r}
-1 & \underline{0} & \underline{0} & \underline{0} \\
\underline{0} & \mathbf{1} & \mathbf{0} & \mathbf{0} \\
\underline{0} & \mathbf{0} & \mathbf{-1} & \mathbf{0} \\
\underline{0} & \mathbf{0} & \mathbf{0} & \mathbf{1} 
\end{array} \right )
\end{array}
\ee 
where
\be
\begin{array}{c c c}
\underline{v}_z \,\equiv \, \left( \begin{array}{c}
0 \\ 1 \\ 0 \end{array} \right) &
\underline{v}_x \,\equiv \, {\sqrt{2} \over 2} \left( \begin{array}{c}
-1 \\ 0 \\ 1 \end{array} \right) &
\underline{v}_y \,\equiv \, -{\sqrt{2} \over 2} \left( \begin{array}{c}
1 \\ 0 \\ 1 \end{array} \right)  \\ \\
\mathbf{J}_z \,\equiv \, \left( \begin{array}{c c c}
1 & 0 & 0 \\ 0 & 0 & 0 \\ 0 & 0 & -1 \end{array} \right) &
\mathbf{J}_x \,\equiv \, {\sqrt{2} \over 2} \left( \begin{array}{c c c}
0 & 1 & 0 \\ 1 & 0 & 1 \\ 0 & 1 & 0 \end{array} \right) &
\mathbf{J}_y \,\equiv \, {\sqrt{2} \over 2} \left( \begin{array}{c c c}
0 & 1 & 0 \\ -1 & 0 & 1 \\ 0 & -1 & 0 \end{array} \right) \\
\end{array}
\ee
The representation generated by these matrices is seen to be finite dimensional,
but not unitary
(analogous to the Dirac representation for spin ${1 \over 2}$ systems).

\setcounter{equation}{0} 
\section{An Extended Poincare Group}
\subsection{Extended Poincare Group Closure}
\tab
The equations presented thus far are valid for internal transformations on
systems.  We will next attempt to minimally expand the group structure to include
space-time translations, in order to develop a group that can be used for global
transformations.  An attempt to only include the momentum operators
does not produced a closed group structure, due to Jacobi relations of the
sort given by $[P_j , [\Gamma ^ 0 , \Gamma ^k] ]$.  The non-vanishing of
this commutator in the Jacobi identity implies a non-vanishing commutator
between $\Gamma ^\mu$ and $P_\nu$, and that this commutator connect
to an operator which then commutes with $\Gamma ^\mu$ to yield a $P_\nu$. 
Since the momentum operators self-commute, this requires the introduction
of at least one additional operator that we will refer to as $\mathcal{G}$.

\subsection{A Closed Set of Extended Poincare Operators}
\tab
The non-vanishing commutators involving the operators $\hat{P}_\mu$ and
$\hat{\mathcal{G}}$ that satisfy the Jacobi identities are given by
\bea
\left [ J_j \, , \, P_k \right] \: = \: i \, \epsilon_{j k m} \, P_m 
\label{JPeqn} \\
\left [ K_j \, , \, P_0 \right] \: = \: -i \, P_j 
\label{KP0eqn} \\
\left [ K_j \, , \, P_k \right] \: = \: -i \, \delta_{j k} \, P_0 
\label{KPeqn} \\
\left [ \Gamma^\mu \, , \, P_\nu \right] \: = \: i \, \delta_\nu ^\mu \, \mathcal{G} 
\label{GamPeqn} \\
\left [ \Gamma^\mu \, , \, \mathcal{G} \right] \: = \: i \, \eta^{\mu \nu} \, P_\nu 
\label{GamGeqn}
\eea
We can construct an extended Poincare group Casimir operator given by
\be
\mathcal{C}_\mu \: \equiv \: \mathcal{G}^2 \,-\, \eta^{\beta \nu} P_\beta P_\nu .
\label{Casimir_mu}
\ee

The extended Lorentz subgroup Casimir operator defined in
Equation \ref{Casimir_Lam} has the following nontrivial commutation
relations:
\be
\left [ \mathcal{G}, C_\Lambda \right ] \: = \: 
i \left ( \Gamma^\mu P_\mu  + P_\mu \Gamma^\mu \right )
\label{GCLam}
\ee
\be
\left [ P_0, C_\Lambda \right ] \: = \:
-i \left ( \Gamma^0 \mathcal{G} + \mathcal{G} \Gamma^0 -
\sum_{j}^{} K_j P_j + P_j K_j    \right )
\label{P0CLam}
\ee
\be
\left [ P_j, C_\Lambda \right ] \: = \:
i \left ( \Gamma^k \mathcal{G} + \mathcal{G} \Gamma^k 
-K_j P_0 -  P_0 K_j  + \epsilon_{jkm} \left(J_k P_m + P_m J_k  \right )  \right )
\label{PClam}
\ee

\subsection{Extended Poincare Group metrics}
\tab 
In general, a group metric can be developed from the adjoint representation
in terms of the structure constants.  The non-vanishing group metric elements
generated by the structure constants of this extended Poincare group are
given by
\bea
\eta^{(P)} _{J_m \, J_n} \: = \: -8 \, \delta_{m,n} &
\eta^{(P)} _{K_m \, K_n} \: = \: +8 \, \delta_{m,n}
\eea
\be
\eta^{(P)} _{\Gamma^\mu \, \Gamma^\nu} \: = \: 8 \, \eta^{\mu \, \nu}
\ee
where $\eta^{\mu \, \nu}$ is the usual Minkowski metric of the Lorentz group.  This
metric is seen to be non-trivially generated by the extended $\Gamma$ algebra.

\subsection{Local Factors}
\tab
For a general ray representation of a quantum mechanical system, 
group transformations on a quantum state vector can introduce
additional phase factors;
\be
U(\underline{b}) \: U(\underline{a}) \: = \: e^{i \zeta (\underline{b};\underline{a})}
U(\underline{\phi}(\underline{b};\underline{a}))
\ee
where $\zeta$ is the local exponent and $\phi$ represents the general
group multiplication element.  The behavior of the local exponent under
group transformations can in general introduce local factors which are c-numbers
into the algebra of the generators.  These local factors
can have physical significance, as in the case of the Galilean group for
non-relativistic transformations\cite{Morrison}.  We will briefly explore
the local factors for this extended Poincare group.

All local factors for this group can be shown to vanish except those in the
following list of commutators: \\
Equations \ref{JKeqn} and \ref{Gam0Gameqn}, where the term on the right
hand sides is shifted by $K_m - \xi_{\Gamma ^m , \Gamma ^0}$; \\
Equations \ref{GamJeqn} and \ref{Gam0Keqn}, where the term on the right
hand sides is shifted by $\Gamma ^m + \xi_{K_m , \Gamma ^0}$; \\
Equation \ref{GamKeqn}, where the term on the right hand side is shifted by
$\Gamma ^0 + \xi_{K_z , \Gamma ^z}$; \\
Equations \ref{JPeqn}, \ref{KP0eqn}, and \ref{GamGeqn}, where the term on the
right hand sides is shifted by $P_m + \xi_{\Gamma_m , \mathcal{G}}$; \\
Equations \ref{KPeqn} and \ref{GamGeqn}, where the term on the right hand sides
is shifted by $-P_0 + \xi_{\Gamma ^0 , \mathcal{G}}$; \\
Equation \ref{GamPeqn}, where the term on the right hand side is shifted by
$\mathcal{G} + \xi_{\Gamma ^0 , P_0}$.\\
The local factors $\xi$ appear in precisely the appropriate manner such that
they can be absorbed into re-definitions of $K_m, \Gamma ^\mu , P_\mu ,$ and
$\mathcal{G}$, eliminating their appearance in the commutation relations.  Henceforth,
we will work with a representation in which all local factors have been eliminated.

\subsection{Unitary Representations of the Extended Poincare Group}
\tab
A natural set of discrete basis states for the extended Poincare algebra
is specified in terms of the group Casimir, the extended Lorentz subgroup
Casimir, and mutually commuting eigenvalues of $\Gamma^0$, $J^2$, and
$J_z$ given by $|\mu^2 , \Lambda, \gamma , J, M>$.  Here $\mu^2$ represents
the eigenvalue of the extended Poincare group Casimir operator, and
the eigenvalue of the extended Lorentz subgroup Casimir
can be expresed in terms of $\Lambda$.  If we are to construct
states that have correspondence with the usual particle states, we need
to examine eigenstates of $\hat{P}_0$.  We have several choices on how to
choose mutually commuting parameters, for instance
$| \mu^2 , p_{(s) 0}, \gamma, J, M>$ or $| p_{(s) 0}, \mathcal{Q}_{(s)}, \gamma, J, M>$. 
We will choose states of the form
\begin{center}
$\left | \sqrt{-\vec{p}_{(s)} \cdot \vec{p}_{(s)}}, \mathcal{Q}_{(s)}=0, \gamma, J, M \right >$
\end{center}
which have vanishing value when operated on by $\mathcal{G}$
as the standard particle states from which we can build unitary representations. 
States of finite momenta are constructed using pure Lorentz boosts upon
the standard state vector:
\be
\left | \underline{u}, \mu, \gamma, J, M \right > \: \equiv \: 
L(\underline{u}) \left | \sqrt{-\vec{p}_{(s)} \cdot \vec{p}_{(s)}}, \mathcal{Q}_{(s)}=0, \gamma, J, M \right >
\label{pstates}
\ee
where $\mu^2=-\vec{p}_{(s)} \cdot \vec{p}_{(s)}$ and $\vec{u} \cdot \vec{u}=-1$ for massive
states.  Unitary transformations
involving general Lorentz transformations $\mathbf{\Lambda}$ give
\be
U(\mathbf{\Lambda})  | \underline{u}, \mu, \gamma, J, M> \:= \:
\sum_{M'=-J}^{J} | \underline{\mathbf{\Lambda}u}, \mu, \gamma, J, M'>
D_{M' , M} ^{(J)} (R_W (\mathbf{\Lambda}, \underline{u}))
\ee
where $R_W (\mathbf{\Lambda}, \underline{u})=
L^{-1}(\underline{\mathbf{\Lambda}u})\mathbf{\Lambda} L(\underline{u})$
is the usual Wigner rotation\cite{Wigner}.  This represents the little
group element for the Lorentz subgroup for the standard state vector
$\vec{p}_{(s)}$.

For more general transformations by group elements $\underline{g}$ conjugate
to the standard state group element $\underline{g}_{(s)}$, the little group of
transformations will leave the standard state group element invariant,
$\mathcal{R} \underline{g}_{(s)}=\underline{g}_{(s)}$.  Since angular momenta and $\Gamma^0$
are hermitian operators, operations involving this subgroup of transformations
will be generalized rotations on the standard state vector that will leave the
standard state group element invariant.  Using the complimentary group
of transformations,  boosted states can be defined:
\be
| \underline{g}, \mu^2, \Lambda, \gamma, J, M> \: \equiv \: 
U(C(\underline{g})) | \underline{g}_{(s)}, \mu^2, \Lambda, \gamma, J, M>
\ee
where $C(\underline{g}) \underline{g}_{(s)} = \underline{g}$.
A unitary representation can then be constructed using
\be
U(\mathcal{M})  | \underline{g}, \mu^2, \Lambda, \gamma, J, M> \:= \:
\sum_{\gamma'=-J}^{J} \sum_{M'=-J}^{J} 
|\mathcal{M} \underline{g}, \mu^2, \Lambda, \gamma', J, M'>
D_{\gamma' M' ; \gamma M} ^{(\Lambda J)} (\mathcal{R} (\mathcal{M}, \underline{g})) ,
\ee
where the little group element satisfies
\be
\mathcal{R} (\mathcal{M}, \underline{g}) \: \equiv \:
C^{-1}(\mathcal{M} \underline{g}) \mathcal{M} C(\underline{g}).
\ee
The group structure of this extended Lorentz and Poincare group will be
explored in subsequent papers.

\subsection{Linear Wave Equation for Single Particle States}
\tab
A primary motivation for this work was to develop general finite
dimensional expressions of the operator $\Gamma^\mu P_\mu$. 
Eigenstates of this operator should give linear operator dispersion
for energy and momenta in a wave equation.  It is straightforward
to calculate the commutators of the various group generators with
this operator:
\be
\left [ J_k, \Gamma^\mu P_\mu \right ] \: = \: 0
\label{JDirac}
\ee
\be
\left [ K_k, \Gamma^\mu P_\mu \right ] \: = \: 0
\label{KDirac}
\ee
\be
\left [ \Gamma^k, \Gamma^\mu P_\mu \right ] \: = \: 
i \Gamma^k \mathcal{G} + i \epsilon_{kmn} J_m P_n - i K_k P_0
\ee
\be
\left [ \Gamma^0, \Gamma^\mu P_\mu \right ] \: = \: 
i \Gamma^0 \mathcal{G} + i \sum_{j} K_j P_j
\label{Gam0Dirac}
\ee
\be
\left [ P_\beta, \Gamma^\mu P_\mu \right ] \: = \: 
-i \mathcal{G} P_\beta
\label{PDirac}
\ee
\be
\left [ \mathcal{G}, \Gamma^\mu P_\mu \right ] \: = \: 
-i \eta^{\beta \nu} P_\beta P_\nu
\label{GDirac}
\ee

Since pure Lorentz transformations commute with the operator
$\Gamma^\mu P_\mu$ from Equations \ref{JDirac} and \ref{KDirac},
we can define a mixed wavefunction using Equation \ref{pstates}
\be
<\mu^2 , \Lambda, \gamma , J, M| \Gamma^\mu P_\mu  
| \underline{u}, \mu, \gamma, J, M> \: = \: \gamma \mu
<\mu^2 , \Lambda, \gamma , J, M| \underline{u}, \mu, \gamma, J, M>
\ee
to obtain a linear wave equation given by
\be
\mathbf{\Gamma}^\mu \hat{P}_\mu \mathbf{\psi}_{\gamma M}^\Lambda (\mu \underline{u}) \: = \:
\gamma \mu \mathbf{\psi}_{\gamma M}^\lambda (\mu \underline{u})
\ee
where
\be
\mathbf{\psi}_{\gamma M}^\Lambda (\mu \underline{u}) \: \equiv \:
<\mu^2 , \Lambda, \gamma , J, M| \underline{u}, \mu, \gamma, J, M>.
\ee
In a configuration space basis, this can be written as
\be
\mathbf{\Gamma}^\mu (-i \partial_\mu) \mathbf{\psi}_{\gamma M}^\Lambda (x) \: = \:
\gamma \mu \mathbf{\psi}_{\gamma M}^\Lambda(x)
\ee
to give the eigenvalue differential equation.

It is clear that since for two non-interacting subsystems the components of the
energy-momentum of clusters are additive, such linear dispersions make explicit clustering
properties more apparent.  It should then be straightforward to include the
kinematic variables of a non-interacting cluster in a purely parametric way
when calculating the dynamics of an off-shell, off-diagonal subsystem.

\subsection{Finite Extended Parameter Transforms}
\tab
We end by examining finite transformations in the extended group parameters. 
To examine transformations on the generators that define the minimal
Poincare extension of the extended Lorentz group, we will express the
group parameter conjugate to the operators $\hat{\Gamma}^\mu$ in terms of
a magnitude and direction, $\omega_\mu \, \equiv \, \omega \, u_\mu$. 
One can directly demonstrate that the commutation relations Equations \ref{GamPeqn}
and \ref{GamGeqn} imply that for time-like ($\vec{u} \cdot \vec{u}=-1$) or
space-like ($\vec{u} \cdot \vec{u}=+1$) transformations
\be
e^{i \omega_\mu \hat{\Gamma}^\mu} \hat{\mathcal{G}} 
e^{-i \omega_\mu \hat{\Gamma}^\mu}\: = \: 
cos(\sqrt{-\vec{\omega} \cdot \vec{\omega}}) \, \hat{\mathcal{G}} \, + \,
{sin(\sqrt{-\vec{\omega} \cdot \vec{\omega}}) \over \sqrt{-\vec{\omega} \cdot \vec{\omega}}}
\omega_\mu \eta^{\mu \nu} \, \hat{P}_\nu
\label{GamGtransf}
\ee
\be
e^{i \omega_\mu \hat{\Gamma}^\mu} \hat{P}_\beta
e^{-i \omega_\mu \hat{\Gamma}^\mu}\: = \: 
\omega_\beta 
{sin(\sqrt{-\vec{\omega} \cdot \vec{\omega}}) \over (\sqrt{-\vec{\omega} \cdot \vec{\omega}})}
\, \hat{\mathcal{G}} \, + \,
\left [ \delta_\beta ^\nu \, + \, 
{{\omega_\beta \omega^\nu}\over \vec{\omega} \cdot \vec{\omega}}(  
cos(\sqrt{-\vec{\omega} \cdot \vec{\omega}}) \, - \, 1) \right ] \, \hat{P}_\nu .
\label{GamPtransf}
\ee
We can see that for eigenstates of $\hat{\Gamma}^0$ such as the finite
dimensional representations discussed previously, the time-like
transformations mix eigenvalues of $\hat{P}_0$ (masses) with eigenvalues
of $\hat{\mathcal{G}}$.  The mass values oscillate under variations in the
parameter $\omega$.  If the group elements $\omega$ are discrete, then the
discrete set of masses will mix into each other and eigenvalues of $\mathcal{G}$
under operations using finite transformations in group elements conjugate to $\Gamma^0$. 
In this sense, the operator $\mathcal{G}$ which is
necessary for group closure can be interpreted as a transverse mass operator. 

Similarly, one can examine the non-trivial transformations by the operators
$\mathcal{G}$ and $\hat{P}_0$.  The operators $\hat{\Gamma}^\mu$ transform
as follows:
\be
e^{i \alpha \hat{\mathcal{G}}} \hat{\Gamma}^\mu e^{-i \alpha \hat{\mathcal{G}}}
\: = \: \hat{\Gamma}^\mu \, + \, \alpha \eta^{\mu \nu} \, \hat{P}_\nu
\label{GGamtransf}
\ee
\be
e^{i a^\beta \hat{P}_\beta} \hat{\Gamma}^\mu e^{-i a^\beta \hat{P}_\beta}
\: = \: \hat{\Gamma}^\mu \, + \,  a^\mu \, \hat{\mathcal{G}}
\label{P0Gamtransf}
\ee
These equations directly indicate that a discrete finite dimensional
representation that can be used to model the behaviors of the usual
particles in quantum field theory cannot be constructed if $\hat{\mathcal{G}}$
has non-vanishing expectation values, or if there is a finite transformation
$\alpha \not= 0$ upon such a state (unless the group parameters 
$\alpha$ are discrete).

\section{Acknowledgements}

The author wishes to acknowledge the support of Elnora Herod and
Penelope Brown during the intermediate periods prior to and after
his Peace Corps service (1984-1988), during which time the bulk of this work was
accomplished.  In addition, the author wishes to recognize the
hospitality of the Department of Physics at the University of Dar
Es Salaam during the three years from 1985-1987 in which a substantial portion of
this work was done.

\end{document}